\documentclass[twocolumn,showpacs,amsmath,amssymb,prd,aps]{revtex4}
\usepackage{graphicx}
\usepackage{epstopdf}
\usepackage[usenames,dvipsnames]{color}
 
\newcommand{\req}[1]{Eq.\,(\ref{#1})}

\begin{document} 
\title{Fugacity and Reheating of Primordial Neutrinos}
\author{Jeremiah Birrell$^{1}$}
\author{Cheng-Tao Yang$^{2,3}$}
\author{Pisin Chen$^{2,3,4}$}
\author{Johann Rafelski$^{5}$}
\affiliation{%
$\quad$\\ 
{$^1$Program in Applied Mathematics,~The University of Arizona, Tucson, Arizona, 85721, USA}}

\affiliation{%
$\quad$\\ 
{$^{2}$Department of Physics and Graduate Institute of Astrophysics, National Taiwan University, Taipei, Taiwan 10617}}
\affiliation{%
$\quad$\\ 
{$^{3}$Leung Center for Cosmology and Particle Astrophysics (LeCosPA),National Taiwan University, Taipei, Taiwan, 10617}}
\affiliation{%
$\quad$\\  
{$^{4}$Kavli Institute for Particle Astrophysics and Cosmology,SLAC National Accelerator Laboratory, Menlo Park, CA 94025, USA}}
\affiliation{%
$\quad$\\ 
{$^{5}$Department of Physics, The University of Arizona, Tucson, Arizona, 85721, USA}}
\date{April 16, 2013}

\begin{abstract}
We clarify in a quantitative way the impact that distinct chemical  $T_c$ and kinetic $T_k$ freeze-out temperatures have on the reduction of the neutrino  fugacity $\Upsilon_\nu$ below equilibrium,   i.e. $\Upsilon_\nu<1$,  and  the increase of the neutrino temperature $T_\nu$  via partial reheating.  We establish the connection between $\Upsilon_\nu$ and $T_k$ via the modified reheating relation $T_\nu(\Upsilon_\nu)/T_\gamma$, where $T_\gamma$ is the temperature of the background radiation.  Our results demonstrate that one must introduce the chemical nonequilibrium parameter, i.e., the fugacity, $\Upsilon_\nu$, as an additional standard cosmological model parameter in the evaluation of CMB fluctuations as its value allows measurement of $T_k$.
\end{abstract}


 \maketitle

\noindent{\bf Introduction:} 
The free-streaming relic neutrino distribution is an important input into structure formation in the universe and the calculation of CMB fluctuations \cite{Lesgourgues:2006nd}.  Recent results from the nine-year WMAP observations, Table 7 of Ref.\cite{WMAP9} and another independent study of BBN~\cite{Steigman:2012ve} favor an effective number of neutrinos at BBN of $N_\nu=3.55^{+0.49}_{-0.48}$ and $N_\nu=3.71^{+0.47}_{-0.45}$, respectively, while the newly released Planck data finds $N_\nu=3.30\pm 0.27$ \cite{Planck}.  However, this fit  produces a 2.5 s.d. tension with direct astrophysical measurements of the Hubble constant. Including priors from SN surveys removes this tension and result in $N_\nu=3.62\pm 0.25$. 

Much work has been done to compute the modification of the neutrino distribution due to two-body interactions between neutrinos and $e^+e^-$ in general relativistic kinetic theory.  Standard electroweak interactions give rise to $N_\nu=3.046$ ~\cite{Lopez:1998aq,Gnedin:1997vn,Mangano:2005cc} while a study of modified neutrino-electron interactions leads to   $N_\nu=3.12$~\cite{Mangano:2006ar}. Current observational data does not have tight enough error bounds to either confirm or show incompleteness of the standard picture of neutrino freeze-out, but the deviation of the central value determined by Planck and others from the computed value suggests that one consider additional factors capable of modifying the neutrino distribution and thereby increasing $N_\nu$.

One  such mechanism is the interaction of neutrinos with collective degrees of freedom in the $e^+e^-$ plasma.  The calculations referenced above assume that the freeze-out process is dominated by two-body scattering, but scattering against collective degrees of freedom is often a significant factor in plasma dynamics.  If such an effect is relevant here, the result would be a lower neutrino kinetic freeze-out temperature $T_k$   and,  as we will show, a larger value of $N_\nu$.  We do not attempt to model the multi-body scattering processes that could lead to a reduction in $T_k$. Rather, we perform a parametric study of the effect that the reduction of $T_k$ has on the neutrino distribution and photon-neutrino reheating ratio within a simple model that emphasizes the physics of interest.  We discover a new effect that accumulates during the temporary reduction of the deceleration parameter $q$ caused by the electron mass, which has a significant effect on the neutrino distribution when $T_k\approx m_e$.

Our study relies on the following two properties of the dynamical evolution of the primordial $e^+,e^-,\nu,\bar \nu, \gamma$ plasma:

\noindent 1) The chemical and kinetic, often called thermal, freeze-out conditions are distinctively different. Once the universe temperature drops below  the chemical freeze-out temperature  $T_c$, there are no reactions that, in a noteworthy fashion, can change the neutrino abundance. We will find that the precise value of $T_c$ is immaterial to the present study as long as $T_c>2m_e$, which is the consensus of all evaluations~\cite{Hannestad:1995rs,Dolgov:2002wy} -- in our own detailed study, we find $T_c\simeq 2.7 m_e$~\cite{forthcoming}. As the universe cools further, at some point neutrinos begin to free-stream~\cite{Birrell:2012gg}.  This is the kinetic freeze-out and $T_k$ is the key parameter in determining the neutrino momentum distribution.  Note that particle creation, which determines $T_c$, is a hard process while exchange of momentum, which determines $T_k$, is soft.  Hence $T_c$ does not have the same complications involving collective phenomena as $T_k$ does. This allows us to treat $T_c$ as being determined by two-body interactions while $T_k$ will be treated as an unknown parameter. 

\noindent 2) There exists a brief time window when the temperature of the universe $T < 2 m_e$, during which the deceleration parameter $q$ drops mildly from that for the pure radiative universe because the electron-positron mass becomes significant and affects the dynamics of the expansion. This lasts until the density of  $e^+e^-$ pairs becomes negligible and the fully radiative expansion resumes at $T<0.15m_e$.

We find that these effects combine to produce a photon-neutrino temperature ratio $T_\gamma/T_\nu$ that is closer to $1$,  a  known `neutrino reheating' effect~\cite{Lopez:1998aq,Gnedin:1997vn,Mangano:2005cc}. However, it has thus far escaped attention  that   reheating of neutrinos can be  accompanied by   a significant underpopulation of neutrino phase space relative to an equilibrium distribution, characterized by a little-known cosmological model parameter which will be called  neutrino fugacity,  $\Upsilon_\nu$. Its significance for neutrino cosmology has been previously recognized \cite{Lattanzi:2005qq}  but is not widely appreciated.  

We  relate  $T_\gamma/T_\nu$ to $\Upsilon_\nu$ and discuss how a significant deviation of $\Upsilon_\nu<1$ alters the neutrino momentum distribution at the time of recombination. Our findings will demonstrate that $\Upsilon_\nu$ should be included among the standard cosmological model parameters governing the experimental CMB observables, especially with regard to neutrino mass constraints. 

\vskip 0.2cm
\noindent{\bf Neutrino  Energy Spectrum and Moments:} 
Consider a particle with degeneracy $g_p$, mass $m$, and chemical freeze-out temperature $T_c$.  Prior to chemical freeze-out  the distribution function has the usual Bose-Einstein or Fermi-Dirac chemical equilibrium form. After chemical but prior to kinetic freeze-out at $T_k$, collisions between particles maintain a maximum entropy distribution, but particle number changing processes are no longer able to occur. The number of particle-antiparticle pairs is therefore conserved. To achieve particle conservation in an environment in which temperature changes, another parameter is needed: $\Upsilon$. The momentum distribution and thus value of $\Upsilon$ is found by maximizing entropy at fixed particle number, antiparticle number, and energy.

The thermal (kinetic) equilibrium momentum distribution functions are applicable down to kinetic freeze-out $T_k$ 
\begin{equation}\label{thermal_eq_dist}
f_{{\rm eq}_{k}}=\frac{g_p}{8\pi^3}\frac{1}{\Upsilon^{-1}e^{ (E-\mu_t)/T}\pm1},
\end{equation}
where  $\Upsilon=1$ for $T>T_c$ and  $\Upsilon\ne 1$ for $T\in [T_c, T_k]$ is the same for both particles and antiparticles. One can also write 
\begin{equation}\label{chem_pot}
\mu=T\ln \Upsilon+\mu_t.
\end{equation}
$\Upsilon$ controls the number of  particle-antiparticle pairs available to fill the phase space. In the absence of (measurable) true chemical potentials $\mu_t$ originated from conserved quantities such as lepton number, $\Upsilon$ is identical to the complete particle fugacity.  Therefore, in this situation an effective chemical potential $\mu$ is derivable from $\Upsilon$ through \req{chem_pot} alone. Chemical non-equilibrium is characterized by $\Upsilon\ne 1$, i.e. for $\mu_t\simeq 0$ but $\mu\ne 0$. For fermions the limit $\Upsilon\gg 1$ approaches a degenerate Fermi distribution, while for $\Upsilon\to 0$ the Boltzmann limit may be used. 

The moments of \req{thermal_eq_dist} define the energy density $\rho$, pressure $P$, number density $n$, and entropy density $s$ of a distribution in kinetic equilibrium \req{thermal_eq_dist}. Standard thermodynamic identities, such as the Gibbs-Durham relations,  remain valid down to  the  kinetic freeze-out stage in the evolution of the universe. In principle, once the neutrinos start free-streaming these relations do not apply anymore. However, in a radiation dominant universe a significant violation occurs only when the mass of the neutrino becomes a significant scale.


\vskip 0.2cm
\noindent{\bf Modeling Universe Expansion:} 
The neutrino oscillations are fast compared to the dynamics of the universe  for  $T \simeq T_k\simeq m_e$. Thus  each neutrino spends one-third of its time in each of the three flavor states and  its kinetic reactivity will be the arithmetic average of the three relevant rates.  Additionally, neutrinos and antineutrinos in a practically matter-antimatter symmetric universe are subject to  the  same dynamics. Finally, at the temperature scale of interest the neutrino masses are negligible. Together, these factors imply a common value of $\Upsilon_\nu$ for all three flavors of neutrinos and their antiparticles. Hence we can collect them all into a single distribution function, with a degeneracy $g_\nu=6$.  

The three relevant dynamical quantities are thus the temperature $T$ of the $\gamma$, $e^+e^-$, $\nu$, $\overline{\nu}$ system, the common neutrino fugacity $\Upsilon_\nu$, and the scale factor $a$ of a  homogeneous, isotropic, and spatially flat universe: 
\begin{equation}
ds^2=dt^2-a(t)^2(dx^2+dy^2+dz^2).
\end{equation}
The dynamics are determined by:\\
a) The Einstein equation. For a flat FRW universe we integrate
\begin{equation}\label{einstein_eq}
H^2\equiv \left(\frac{\dot{a}}{a}\right)^2=\frac{\rho}{3 M_p^2}, \hspace{2mm} M_p=1/\sqrt{8\pi G},
\end{equation}
where $\rho$ is the total energy density of photons, $e^+e^-$, and neutrinos. \\
b) Divergence freedom of the total stress-energy tensor (which is implicit in the Einstein equation). Using the Gibbs-Durham relations, we find that divergence freedom of the 
stress-energy tensor, 
\begin{equation}\label{stress_energy_eq}
\nabla_\mu \mathcal{T}^{\mu 0}=\dot{\rho}+3\left(\rho+P\right)\frac{\dot{a}}{a}=0,
\end{equation}
is equivalent to
\begin{align}\label{S_n_eq}
T\frac{d}{dt}(a^3s)&+\sum_i\mu_i\frac{d}{dt}(a^3n_i) =0.
\end{align}
c) A dynamical equation governing the neutrino-number-changing  process, 
\begin{equation}\label{chem_eq_process}
e^++e^-\longleftrightarrow \nu_e+\overline{\nu}_e,
\end{equation}
which reads  
\begin{equation}\label{dyn_upsilon}
\dot{\Upsilon}_\nu+3\Upsilon_\nu\!\left(\!H+\frac{\dot{T}}{T}\!\right)\!\!\frac{n_\nu/\Upsilon_\nu}{\partial n_\nu/\partial \Upsilon_\nu}=\left(\Upsilon_e^2-\Upsilon_\nu^2\right)\frac{1}{\tau_e}.
\end{equation}
Here $\Upsilon_e^2\to 1$. For further details of the derivation of \req{dyn_upsilon} see  Eq.(7.290) in Ref.\cite{Inga}, of which this is a slight modification, as well as Refs.\cite{Kolb:1990vq,Bernstein}.  

Neutrino oscillations allow us to collect all three neutrino flavors into a single distribution for which the dominant reaction strength is that in \req{chem_eq_process}, which is the reaction most  responsible for chemical equilibrium as it is the only rate that contains charged current (exchange of $W^\pm$), with small contributions by the neutral current ($Z^0$) mediated $e^+e^-$ annihilation, which also feeds into $\nu_\mu$ and $\nu_\tau$.  Since each of the three neutrino flavors can be fed in $e^+e^-$ annihilation the total relaxation time is 
\begin{equation}\label{tau_all}
 \frac{1}{\tau}=\frac{1}{\tau_e}+\frac{1}{\tau_\mu}+\frac{1}{\tau_\tau}\simeq \frac{1}{\tau_e},
\end{equation}
where the electron rate dominates within the precision required in our approach.

The  rate constant $\tau_e(T)$  is obtained by conventional kinetic methods described in Ref.\cite{Inga}; see Ref.\cite{forthcoming} for more details. Our electroweak amplitudes and invariant rates agree with those presented in Refs.\cite{Hannestad:1995rs,Dolgov:2002wy}. We obtain, to better than 5\% precision in Boltzmann approximation, the result
\begin{equation}\label{tau_T}
\frac{1}{\tau_e}\simeq\frac{2^{-5}G_F^2T^8}{1+(2m_e/T)^2e^{0.29(2m_e/T-16)}}\left(\frac{\partial n_\nu}{\partial \Upsilon_\nu}\right)^{-1}\!\!\!\!,
\end{equation}
where $G_F$ is the Fermi coupling constant.  The neutrino degeneracy in the last normalizing term in \req{tau_T} is $g_\nu=2\times 3_f$. Note that the flavor factor $3_f$ averages the rate constant shown in \req{tau_all} and thus the effective rate of chemical equilibration is 1/3 of that for $\nu_e$ alone in absence of neutrino oscillations.

This kinetic theory model describes a smooth chemical freeze-out process, as is inherent in \req{dyn_upsilon}. This is in contrast to the instantaneous transition approximation, in which at some temperature  $T_c$ it is assumed that all neutrino number changing processes cease completely.  Under this `sudden' approximation, the two terms in \req{S_n_eq} are separately  zero and and \req{S_n_eq} and \req{dyn_upsilon} are replaced by the independent conservation of comoving entropy and neutrino number. 

In the more precise `smooth' model of \req{dyn_upsilon}, comoving entropy is not exactly conserved. In general, particle number changing processes occurring outside chemical equilibrium, i.e. $\Upsilon\ne 1$, are entropy generating processes. We verified  that in our case in ultrarelativistic plasma, the conversion of $e^+e^-$ pairs into neutrino pairs is practically entropy conserving. This explains why under  the entropy conserving approximation, $\Upsilon_\nu(T)$ remains accurate to within a few percent, a point we will  demonstrate explicitly. Thus, in principle there is no need to model the chemical freeze-out in detail for the purpose of the discussion carried out  below and we are justified in assuming exact conservation of entropy when convenient. This would of course be anyone's first guess but considering the issues of reheating we address, we have convinced ourselves by modeling the smooth chemical freeze-out that this is an accurate model assumption. 

\vskip 0.2cm

\begin{figure} 
\centerline{\includegraphics[height=6cm]{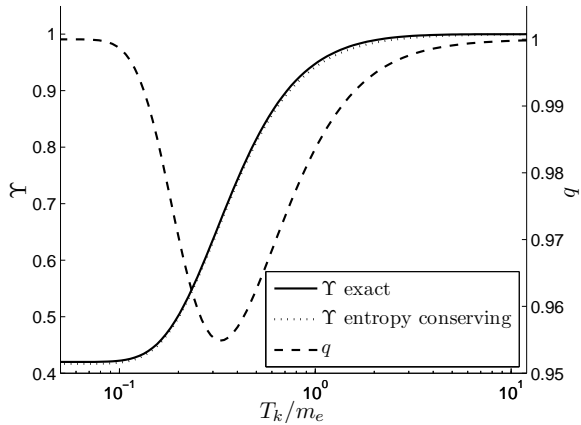}}
\caption{Neutrino fugacity $\Upsilon_\nu$ as a function of neutrino kinetic freeze-out temperature $T_k/m_e$. The solid line corresponds to the exact numerical solution for $\Upsilon_\nu$ and the dot-dashed line closely following it is the entropy conserving freeze-out approximation; see the text for details.  Dashed line (scale on right) is the deceleration parameter.}\label{fig:Upsilon_q}
 \end{figure}

\vskip 0.2cm
\noindent{\bf Effect of Electron Mass:} 
If all relevant particles have negligible mass then both entropy and number density scale with $T^3$, hence the solutions to \req{S_n_eq}  and \req{dyn_upsilon} are simply
\begin{equation}\label{radiative_sol}
\Upsilon_\nu=1, \hspace{2mm} T\propto \frac{1}{a}.
\end{equation}
However, the proximity of $T_c$ and $T_k$  to the electron mass modifies the situation. As the temperature passes through the electron mass, the  deceleration parameter in the expansion of the Universe, 
\begin{equation}\label{q_decel}
q=-\frac{a\ddot a}{\dot a^2}.
\end{equation}
displays a small deviation from  the pure radiation dominant value of  $q=1$, decreasing slightly towards the matter dominant value of $q=1/2$, indicating a temporary period of matter relevance.  It is this drop in $q$ that causes $\Upsilon_\nu$ to decrease and the neutrino phase space to become underpopulated.   Though the magnitude of the drop in $q$ (right scale in  figure~\ref{fig:Upsilon_q}) is not large, the effect on $\Upsilon_\nu$ is cumulative.  

We obtain a quantitative measure of this underpopulation effect by numerically solving \req{einstein_eq}, (\ref{S_n_eq}), and (\ref{dyn_upsilon}) to obtain $\Upsilon_\nu(T_k)$ for each choice of the neutrino kinetic freeze-out temperature $T_k$, which we treat as a free parameter.  We begin at a temperature high enough, e.g., $T=10m_e$, for the universe to be fully radiation dominant and neutrinos to be in chemical equilibrium with $\Upsilon_\nu=1$, and integrate across $T_c$ until $T=T_k$, at which point the neutrinos become free-streaming.  The result is shown as a solid line in figure \ref{fig:Upsilon_q}.  We also show the result of the entropy conserving  chemical freeze-out approximation as a dot-dashed line, which closely parallels the solid line. In the entropy conserving approximation, the calculation is insensitive to the precise value used for the neutrino chemical freeze-out temperature $T_c$, as long  as it  is several times the electron mass, i.e., larger than  the temperature where $\Upsilon_\nu$ begins to deviate from $1$. This is so since neutrino chemical freeze-out occurs in the nearly exact  radiation dominant era of the universe and prior to the temperature where the influence of the electron mass on the universe dynamics takes effect. In this domain the neutrino distribution evolves as in \req{radiative_sol} irrespective of whether or not it is in chemical equilibrium. The difference between these two calculations is less than $1\%$, so for the purposes of this work, the approximate model is valid.

Delaying the thermal freeze-out condition until it is close to or below the electron mass results in neutrinos acquiring energy liberated by $e^+e^-$ annihilation via scattering and thus are reheated, along with photons, down to $T_k$ . Once thermal freeze-out occurs, only photons are reheated and from that point on $T_\gamma>T_\nu$. At $T_k$ neutrinos begin to free stream and the neutrino entropy is conserved independently of the entropy in photons and $e^+e^-$, as shown in Ref.\cite{Birrell:2012gg}. We make the approximation that the total comoving entropy is exactly conserved, hence whatever entropy remains in $e^+e^-$ at $T_k$ goes solely into reheating the photon temperature.  

The calculation of the resulting photon-neutrino temperature ratio  $T_\gamma/T_\nu$ is the same as that used to derive the traditional reheating ratio of $\left(11/4\right)^{1/3}$, only  that here one starts  with less comoving entropy in $e^+e^-$ since by the time $T=T_k$, some of the entropy has already been transferred to neutrinos. We carry out this procedure numerically and find a simple power-law relation to better than $1\%$ between the reheating ratio $T_\gamma/T_\nu$ and $\Upsilon_\nu$: 
\begin{equation}\label{reheat_ratio}
\Upsilon_\nu=0.420\left(\frac{T_\gamma}{T_\nu}\right)^{2.57}.
\end{equation}
Note that for $\Upsilon_\nu\to 1$, we obtain the standard reheating ratio $T_\gamma/T_\nu=(11/4)^{1/3}$, as expected.  Reducing $T_k$ leads to a reduction of $T_\gamma/T_\nu$ compared to standard reheating, and introduces a significant deviation  of $\Upsilon_\nu$ from unity, an effect which has not been recognized before.

\vskip 0.2cm
\noindent{\bf Momentum distribution at recombination:} 
In Ref.\cite{Birrell:2012gg} we consider and solve the Einstein-Vlasov equation for the free-streaming form of the neutrino momentum distribution after kinetic freeze-out.  Using the fact that $m_\nu\ll T_k$, the result is
\begin{equation}\label{neutrino_dist}
f_\nu(t,p)=\frac{g_\nu}{8\pi^3}\frac{1}{\Upsilon_\nu^{-1}e^{p/T_\nu}+ 1},
\end{equation}
where the neutrino temperature $T_\nu$ is obtained  by solving \req{reheat_ratio}. This is a nonequilibrium distribution for two reasons:\\ 
a) The appearance of  an  effective chemical potential \req{chem_pot}: $\mu_\nu=T_\nu\ln \Upsilon_\nu$.\\
b) The absence of the neutrino mass in the exponential means that when the temperature approaches and drops below $m_\nu$,  the distribution is no longer of the kinetic equilibrium form. 

Depending on the ratio of neutrino mass to recombination temperature, one or both of these effects may be significant. From \req{neutrino_dist} we see that the number density is calculated as if the neutrino were massless, but to compute the energy density or momentum one weights the energy or momentum of a massive particle with an effectively massless distribution function at a temperature $T_\nu$. A fugacity $\Upsilon_\nu\neq 1$ together with the modified reheating ratio $T_\gamma/T_\nu(\Upsilon_\nu)$ imply that the effect of the neutrino mass being close to the recombination temperature is altered from the standard result when $T_k$ is reduced.

\vskip 0.2cm
\noindent{\bf Interpretation of Planck  Results~\cite{Planck}:}  
To explain a value  $N_\nu=3.62\pm0.25$ at ion-electron recombination, we find $T_k=0.19^{+0.07}_{-0.04}$ MeV and 
$\Upsilon_\nu=0.71^{+0.1}_{-0.08}$. The value of effective chemical potential $\mu_\nu$  at recombination, \req{chem_pot}, follows, noting that $\Upsilon_\nu$  is preserved in the free-streaming neutrino momentum distribution and hence it continues to be present at the time of recombination. We find $\mu_\nu\simeq -0.086$ eV. 

The above estimates explicitly assume that the effect we present here is the only one available to explain the Planck results (with priors) for $N_\nu$. Furthermore, we reinterpret the fitted $N_\nu$ rather than performing a fit in which $\Upsilon_\nu\ne 1$ and the reheating constraint \req{reheat_ratio} is included. The neutrino momentum distribution that would enter such a fit would be a background gas with a smaller neutrino number density and greater momentum than currently assumed. Incorporating $\Upsilon_\nu$ together with the modified reheating ratio, \req{reheat_ratio}, into fits of CMB data has the potential to constrain the magnitude of the neutrino mass with greater consistency. 

To interpret the literature value $N_\nu=3.046$, we require  $T_k^{(0)}=0.806$ MeV,  which leads to an effective chemical potential $\mu_\nu=T_\nu\ln \Upsilon_\nu\simeq -2.2\times 10^{-2} T_\nu$.  The value  $T_k$ we find to interpret  a value $N_\nu\simeq 3$ is very sensitive to the method of achieving the kinetic decoupling; it corresponds to $T_k^{\rm col}=0.7864$ MeV obtained by computing directly the two-body collision freeze-out condition~\cite{forthcoming}. The theoretical understanding of how a reduction from the lowest order standard model two-body scattering value $T_k^{\rm col}\simeq 0.8$ MeV to $T_k\simeq 0.2$ MeV is possible, is a question requiring further study.

\vskip 0.2cm
\noindent{\bf Summary and Conclusions:}  
We have derived the dependence of the background cosmic neutrino distribution and reheating ratio on the neutrino thermal freeze-out temperature $T_k$.  The drop  in the deceleration parameter of the universe when the electron mass becomes a relevant scale, $T\simeq m_e$, combined with distinct neutrino freeze-out temperatures that satisfy the condition $T_c>m_e> T_k$, where $T_c$ is the neutrino chemical freeze-out temperature, result in a deviation of the neutrino distribution from equilibrium. This effect is described by a primordial neutrino fugacity $\Upsilon_\nu$ different from unity.

The standard cosmological model determines neutrino freeze-out by two-body microscopic scattering~\cite{Mangano:2005cc,Gnedin:1997vn}.  However, currently a value of $3.6> N_\nu> 3$ is favored by BBN related observations~\cite{WMAP9,Steigman:2012ve} and the recent  Planck fit~\cite{Planck} with priors. The proposed modification of $T_k$ from a two-body scattering value is motivated by the mechanism of neutrino scattering against collective degrees of freedom in the $e^+e^-$ plasma, but in this work $T_k$ was left as a free-parameter. 

We determined the magnitude of the fugacity $\Upsilon_\nu$ as a function of  $T_k$, see figure \ref{fig:Upsilon_q}.  We established a modified reheating relation between the current photon and neutrino temperatures required by  $\Upsilon_\nu\ne 1$, \req{reheat_ratio},   and thus demonstrated  a possible large modification of the primordial neutrino momentum distribution. $\Upsilon_\nu$  should thus serve as a cosmological model parameter, hitherto presumed to be close to unity, in the evaluation of CMB fluctuations. 

A value well above $N_\nu \simeq 3$  would support a lower kinetic freeze-out temperature $T_k$ and hence a larger effective chemical potential in absolute value, up to $\mu_\nu\simeq-0.87\,T_\nu$, maybe required. The neutrino distribution evolves after freeze-out as a free-streaming gas~\cite{Birrell:2012gg} and this preserves the value of $\Upsilon_\nu$, implying that neutrinos have a temperature dependent effective chemical potential $\mu_\nu=T_\nu\ln \Upsilon_\nu$ after freeze-out. One must note that  at the recombination temperature $T_r=0.25 {\rm eV}$,  this effective chemical potential is on the order of the neutrino mass limit and is therefore non-negligible.

{\bf Acknowledgments}
JR is supported by a grant from the U.S. Department of Energy, DE-FG02-04ER41318.
JB is supported by the Department of Defense (DoD) through the National Defense Science \& Engineering Graduate Fellowship (NDSEG) Program.
CTY and PC are supported by Taiwan National Science Council under Project No. NSC 101-2923-M-002-006-MY3 and 101-2628-M-002-006- and by Taiwan's National Center for Theoretical Sciences (NCTS). PC is in addition supported by US Department of Energy under Contract No. DE-AC03-76SF00515.



\end{document}